# Electrical double layer and capacitance of TiO$_2$ electrolyte interfaces from first principles simulations

Chunyi Zhang[1], Marcos Calegari Andrade[2], Zachary K. Goldsmith [1], Abhinav S. Raman[1], Yifan Li[1], Pablo Piaggi[1], Xifan Wu[3], Roberto Car[1]*, and Annabella Selloni[1]*

[1] *Department of Chemistry, Princeton University, Princeton, New Jersey 08544, USA*

[2] *Materials Science Division, Lawrence Livermore National Laboratory, Livermore, California 94550, USA*

[3]*Department of Physics, Temple University, Philadelphia, Pennsylvania 19122, USA*

*Email: rcar@princeton.edu, aselloni@princeton.edu*

**ABSTRACT:**

The electrical double layer (EDL) at aqueous solution-metal oxide interfaces critically affects many fundamental processes in electrochemistry, geology and biology, yet understanding its microscopic structure is challenging for both theory and experiments. Here we employ *ab initio*-based machine learning potentials including long-range electrostatics in large-scale atomistic simulations of the EDL at the TiO$_2$-electrolyte interface. Our simulations provide a molecular-scale picture of the EDL that demonstrates the limitations of standard mean-field models. We further develop a method to accurately calculate the electrostatic potential drop at the interface. The computed capacitance originating from the adsorbed charges and the potential drop agrees with experiments, supporting the reliability of our description of the EDL. The larger interfacial capacitance of basic relative to acidic solutions originates from the higher affinity of the cations for the oxide surface and gives rise to distinct charging mechanisms on negative and positive surfaces.

**INTRODUCTION**

When a metal oxide is interfaced with an aqueous electrolyte with a pH different from its pH point of zero proton charge (pH$_{PZC}$), a net charge forms on the surface[1-4]: at pH < pH$_{PZC}$, the surface adsorbs protons from the electrolyte and becomes positively charged, while at pH > pH$_{PZC}$ it adsorbs hydroxide ions and becomes negatively charged. Electrolyte ions of opposite charges are then drawn closer to the interface to balance the surface charge. This electrostatic attraction, competing with thermal fluctuations and (de)solvation thermochemistry, leads to inhomogeneous ion distributions near the interface[1-4]. Additionally, adjacent water molecules reorient in response to the surface and ionic charges[1-4]. The combination of the charged surface, adjacent ions, and neighboring water molecules constitutes what is known as the electrical double layer (EDL). The EDL governs the chemical reactivity and physical properties of the interface and is crucial in diverse environmental, biological, colloidal, and electrochemical processes[5-8].

The EDL is commonly described using the Gouy-Chapman-Stern (GCS) model, which provides a mean-field picture of the solid-electrolyte interface[9,10]. However, as our understanding of these interfaces has advanced through theory and experiment, the inherent limitations of the GCS model have come to the forefront[3-6]. Notably, GCS assumes a uniformly charged, laterally homogeneous surface. Yet, numerous studies highlight the impact of surface structure on interfacial properties[11]. Similarly, GCS treats water as a homogeneous dielectric continuum, overlooking its molecular character. However, it is well established that the properties of the first few interfacial water layers deviate considerably from those of bulk water[11]. Moreover, increasing experimental evidence underscores the inadequacy of classical mean-field descriptions for dynamic events such as electron transfer and chemical reactions[3,5-8,12,13]. The gaps in the molecular-scale understanding of the EDL have hindered theoretical advances and practical applications in electrochemical systems, energy storage, and interface-driven reactions.

Although experimental surface-sensitive techniques are becoming increasingly available to investigate the structure and dynamics of solid-liquid interfaces[3,14,15], the EDL's microscopic properties are still difficult to probe experimentally. Many studies have thus relied on computer simulations to complement experimental observations and obtain atomistic information. In this regard, (reactive) force field simulations and multiscale modeling have been widely used and provided important insights[10,14], but are generally not accurate enough for describing reactive processes such as water dissociation and proton transfer at aqueous interfaces. In contrast, density functional theory (DFT)-based *ab initio* molecular dynamics (AIMD) can in principle provide more accurate information on both atomistic details of the EDL and on macroscopic observables[16,17]. However, the computational cost of AIMD simulations makes them practical only for system sizes (hundreds of atoms) and simulation times (tens of picoseconds) that are often insufficient to characterize the EDL and pH-dependent surface chemistry[17,18]. For example, AIMD provided estimates of Helmholtz capacitances at oxide-electrolyte interfaces[19,20], but could not predict equilibrium ion distributions in the EDL, for which longer simulation times and larger cells would be needed.

In the past few years, machine-learned potentials (MLPs)[21-25] have emerged as a viable approach to enable large-scale *ab initio*-level simulations of bulk systems, interfaces, and reactive processes[18,26-31]. Of such MLP methods, the Deep Potential (DP) scheme[22,23] has been successfully applied to model bulk aqueous electrolytes[29,32,33] and water-oxide interfaces[18,30,31]. Moreover, this scheme has recently been extended to include long-range electrostatic interactions[34], a component missing in standard MLPs but found to be important for simulating the acid-base chemistry in water[35].

In this work, we use the deep potential long-range (DPLR) method[34] to conduct large-scale molecular dynamics simulations of the anatase $TiO_2$ (101) surface in contact with electrolyte solutions at various pHs. $TiO_2$ is a typical functional oxide and one of the most widely used materials in (photo-)electrochemistry[36-38]. Our simulations provide a comprehensive molecular-scale picture of the EDL at $TiO_2$ interfaces and show the occurrence of distinct microscopic surface charging mechanisms for negative and positive oxide surfaces, demonstrating the limitations of the GCS model. We also developed a method to calculate the electrostatic potential profile with *ab initio* DFT accuracy from the ion distributions and the electron centers provided by the DPLR and Deep Wannier (DW)[39] neural network models. This enables us to evaluate the interfacial capacitance due to the adsorption of protons or hydroxide ions and the corresponding counterions, a macroscopic property measurable from experiments that is highly reflective of the EDL's nature[19,20,40]. The computed capacitances agree with experimental results[40], underscoring the reliability of our *ab initio*-based, machine-learned description of the EDL.

## SIMULATIONS

The DPLR and DW neural network models were trained on a comprehensive set of DFT-SCAN[41] data collected using an active learning approach[23] (see Supplementary Section 1 for training, validation, and MD simulation details). Standard DP (hereafter denoted DPSR) fails to properly describe oxide-electrolyte interfaces, because, in the absence of long-range interactions, charge accumulation at the interface due to surface charging and/or EDL formation may result in charge neutrality violation in the bulk of an electrolytic solution. DPLR remedies this deficiency because the electrostatic energy penalizes bulk charging (Supplementary Section 2). In the absence of dissolved ions, as in the case of the anatase (101) interface with neat water, DPLR gives results that agree with DPSR (Supplementary Section 1.6).

DPLR was used to perform MD simulations on periodically repeated systems consisting of a five-layer (3 × 9) anatase (101) slab in contact with a 67 Å thick aqueous electrolyte (Fig. 1a). We focused on three experimentally relevant solutions[40], namely neutral 0.4 M NaCl (NaCl$_{(aq)}$, composed of 2376 H$_2$O molecules and 18 NaCl), acidic 0.4 M NaCl + 0.2 M HCl (with 10 HCl added to the neutral solution), and basic 0.4 M NaCl + 0.2 M NaOH (with 10 NaOH added to the neutral solution). In the electrolyte, NaCl serves as a background salt, and adding 0.2M NaOH or HCl changes the pH from a nominal value of 7 to 13.3 or 0.7. We note that with 2376 water molecules, adding a single H$^+$ or OH$^-$ ion yields a pH of 1.6 or 12.4, so pH values 1.6 < pH < 12.4 are not accessible to our simulations. The magnified view of the interface (Fig. 1b) shows that the TiO$_2$ surface exposes five-fold coordinated titanium (Ti$_{5c}$) and two-fold coordinated oxygen (O$_{2c}$) atoms (108 Ti$_{5c}$ and 108 O$_{2c}$ sites in total, considering the two surfaces of the slab). The undercoordinated Ti$_{5c}$ atoms act as (Lewis) acid sites for the adsorption of water or OH$^-$ ions, while the O$_{2c}$ atoms act as Brønsted bases which can accept a hydrogen bond or a H$^+$ from water. Water dissociation thus results in protons on O$_{2c}$ sites (bridging hydroxyls) and OH$^-$ groups on Ti$_{5c}$ sites (terminal hydroxyls), as depicted in Fig. 1b.

## RESULTS AND DISCUSSION

### Structure of the EDL

For the TiO$_2$-NaCl$_{(aq)}$ interface, the simulations show an equal number of adsorbed H$^+$ ($N_{H+}$) and OH$^-$ ($N_{OH-}$) species, originating from the dissociation of adsorbed water molecules.

Consequently, the surface charge density (black line in Fig. 1c) is close to zero. This suggests that the pH of the neutral NaCl$_{(aq)}$ is approximately equal to the pH$_{PZC}$ of anatase, a result consistent with the experimental pH$_{PZC}$ of anatase being around 6 (±1) [42-44]. To support this inference, we performed enhanced sampling DPLR simulations to explicitly evaluate the pH$_{PZC}$ of the anatase (101)-water interface (Supplementary Section 3). We obtained a pH$_{PZC}$ of 7.0±0.1, a value within the experimental range.

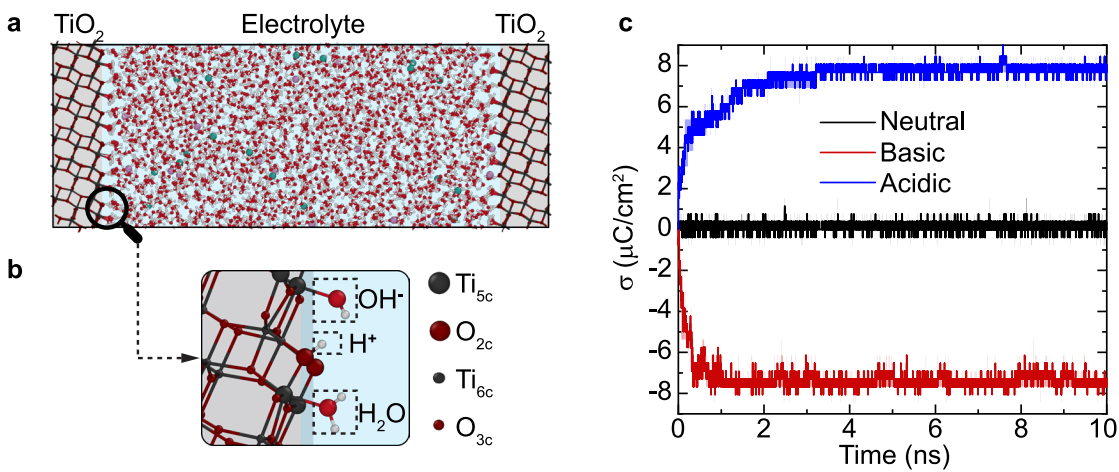

**Fig. 1 | Surface charging. a,** Supercell of the anatase (101)-electrolyte interface employed in the DPLR simulations. The TiO$_2$ regions are shaded in grey, while the electrolyte region is shaded in blue. **b,** Magnified view of the interface showing surface undercoordinated Ti$_{5c}$ and O$_{2c}$ sites and adsorbed H$_2$O, H$^+$ and OH$^-$ at these sites. For visual clarity, only relevant atoms are shown. **c,** Surface charge density $\sigma = e(N_{H+} - N_{OH-})/S$, where $S$ is the surface area, as a function of simulation time for anatase (101) in contact with three different types of electrolytes. All results and error bars (shaded areas) were derived from simulations using two independent DPLR models.

Interfaces of TiO$_2$ with acidic and basic solutions were generated starting from an equilibrated configuration of TiO$_2$-NaCl$_{(aq)}$, and subsequently adding 0.2 M HCl or NaOH at random positions within the electrolyte solution. All additional H$^+$ (or OH$^-$) ions were gradually adsorbed on the surface within 3 ns, leading to a positively (or negatively) charged surface (Fig. 1c). Averaging over 3-10 ns, we obtained surface charge densities $\sigma_a = 7.69 \pm 0.04 \ \mu C/cm^2$ and $\sigma_b = -7.54 \pm 0.13 \ \mu C/cm^2$ at the interfaces with the acidic and basic electrolyte, respectively.

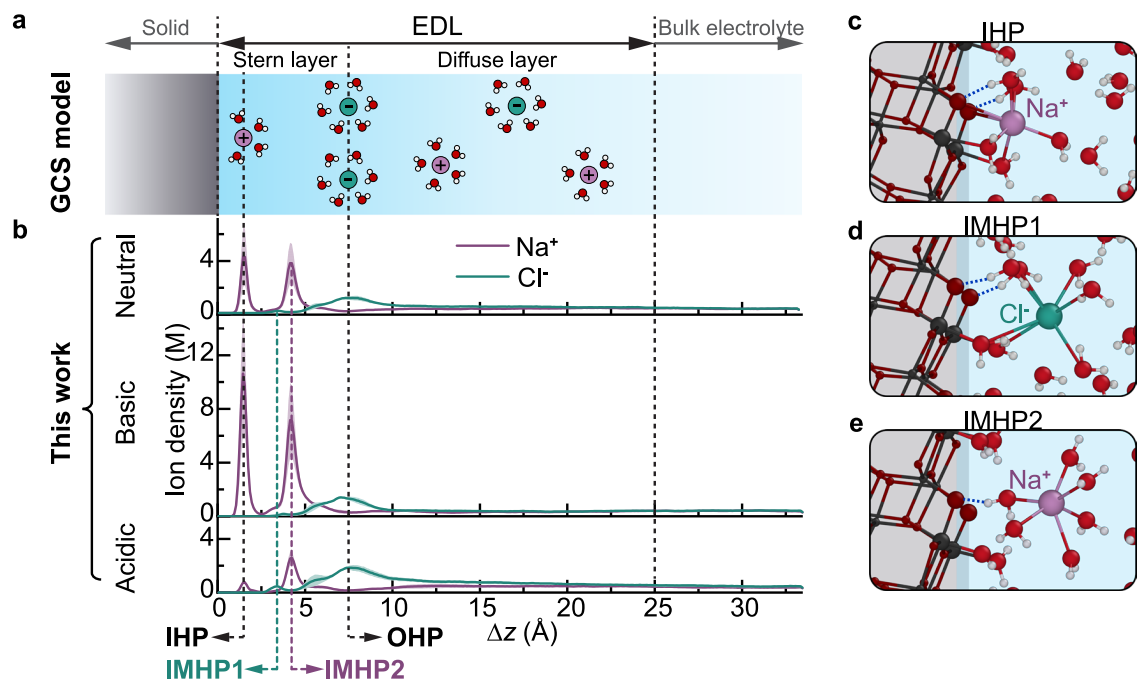

**Fig. 2 | Ion distribution. a,** Schematic of the Gouy-Chapman-Stern (GCS) model of the electric double layer (EDL): the outer Helmholtz plane (OHP) separates the EDL into Stern and diffuse layers. Within the Stern layer, the inner Helmholtz plane (IHP) is defined by the distance at which ions specifically adsorb. **b,** Ion density distributions as functions of distance $\Delta z = z - z_{\text{surface}}$ from the solid surface, obtained from DPLR simulations of anatase (101) in contact with different electrolytes; the position of the solid surface, $z_{\text{surface}}$, corresponds to the average position of the $O_{2c}$ sites. Besides the IHP predicted by the GCS model, these simulations reveal one or two additional ionic density peaks within the Stern layer, which we define as the intermediate Helmholtz planes (IMHPs). The computed ion densities are averaged over the two interfaces in the supercell. All results and error bars were derived from simulations using two independent DPLR models. **c-e,** Illustrative snapshots of the microscopic structures of IHP, IMHP1, and IMHP2. Color code: purple ($Na^+$), green ($Cl^-$), white (H), light red (O in water or water ions), dark red (O in $TiO_2$), and grey (Ti). Lines between ions and neighboring oxygens indicate distances smaller than the ionic hydration shell radius. Dashed blue lines indicate hydrogen bonds between water and surface $O_{2c}$. For visualization purposes, only the most relevant atoms are shown.

Upon surface charging, an EDL forms to compensate the surface charge[1-4]. The electrolyte ion distribution predicted by the GCS model is illustrated in Fig. 2a. The aqueous ions form discrete (Helmholtz) planes of ion adsorption, and the outer Helmholtz plane (OHP) separates the Stern layer from a diffuse layer where the ions follow the Poisson-Boltzmann distribution. The counter-ion and co-ion densities converge to an equal value in the EDL's tail, indicating that the electrolyte recovers its bulk-like behavior. Within the Stern layer, the inner Helmholtz plane (IHP) is formed by ions that are specifically adsorbed at the surface and lack a complete hydration shell (Fig. 2a).

The simulations reveal that the ion distribution at the electrolyte-$TiO_2$ interface is more complex than suggested by the GCS model and by previous force field simulations[14] (Fig. 2b). Specifically, within the Stern layer, we identify not just a single ionic peak as in the GCS model, but two or three distinct peaks: a first $Na^+$ peak at 1.5 Å from the surface, followed by a small $Cl^-$ peak at 3.4 Å (mostly seen in the acidic solution), and another $Na^+$ peak at 4.2 Å. As depicted in Fig. 2c, $Na^+$ ions forming the first peak are coordinated to surface $O_{2C}$ atoms and exhibit incomplete hydration shells. Thus, these ions can be identified as inner-sphere surface complexes[14] and the corresponding peak as the IHP. In contrast, the $Cl^-$ ions forming the second peak (Fig. 2 d) and the $Na^+$ ions contributing to the third peak (Fig. 2 e) exhibit complete hydration shells. However, some water molecules in these shells are either adsorbed on $Ti_{5c}$ atoms or form strong hydrogen bonds with surface $O_{2c}$ atoms, so that their diffusion is very small[45]. These ions can be identified as outer-sphere surface complexes[46]. Their peaks are absent in the GCS model, which treats water as a homogeneous dielectric continuum and overlooks the surface structure. Since the locations of the second and third ionic peaks fall between the IHP and the OHP, we designate the corresponding ionic layers as intermediate Helmholtz planes (IMHPs). On the other hand, outside the OHP, our simulations agree with the GCS model, showing that within the diffuse layer the $Cl^-$ density decreases and the $Na^+$ density increases with increasing distance from the OHP until the two densities become identical within the error bars of the simulation.

The GCS model predicts that the EDL only forms when a net (electronic or protonic) surface charge is present. In contrast, our results show that the charge-neutral $TiO_2$-$NaCl_{(aq)}$ interface also exhibits an EDL (Fig. 2b) because, even when the overall surface charge is zero, local charges remain imbalanced. Specifically, the outermost oxide surface layer is composed of electronegative $O_{2c}$ atoms, which favor the adsorption of $Na^+$ ions and, in combination with the water molecules adsorbed or hydrogen bonded to the surface, determine the ionic peak positions (Fig. 2c-e). Therefore, altering the surface charge density modifies the peak intensities of the ion distribution but does not appreciably change their relative positions (Fig. 2b). In particular, regardless of the surface charge, anions dominate at the OHP and in the diffuse layer to compensate cations in the IMHP2 – two important features missed by the GCS model.

The water molecules within the EDL exhibit distribution and orientation patterns which, as the ion distributions, are mostly determined by the surface structure. Our DPLR simulations show that the water distribution and orientation undergo subtle changes in response to both the surface charge and the adjacent ions (Supplementary Section 4).

**Differential capacitance**

The microscopic structure of the EDL determines the differential capacitance $C$ of the interface, which is defined as the first-order derivative of the surface charge density $\sigma$ with respect to the electrostatic potential drop $\psi$, namely $C = \frac{d\sigma}{d\psi}$. Interestingly, experiments found that, at equivalent magnitudes of charge densities, negatively charged oxide interfaces possess a higher capacitance than their positively charged counterparts (at variance with GCS's prediction)[40,47]. This suggests the EDL's enhanced capability in screening negative surface charges compared to positive ones.

To compute $C$, we used the finite difference expression $C \approx \frac{\Delta\sigma}{\Delta\psi}$, where $\Delta$ represents the deviation of a charged interface from a neutral reference interface. While $\sigma$ is readily available in simulations (Fig. 1c), $\psi$ depends not only on the distribution of the ions but also on that of valence electrons, which is not available in simulations based on force fields. However, $\psi$ is an average property that only requires knowledge of the plane-averaged density profile of the electrons along $z$, which can be estimated accurately from the positions of the Wannier centroids, provided by the DW[31] model, and from the average spread of the associated electron distributions, provided by DFT calculations on smaller systems. The density profile of the total charge (ions + valence electrons) is obtained by adding the plane-averaged ion density profile to the valence electron density profile. Then, $\phi(z)$, the electrostatic potential profile, is calculated by solving a one-dimensional Poisson's equation. As shown in Supplementary Section 5, this procedure is remarkably accurate when compared to DFT calculations on reference systems. The average potential profile calculated in this way for the $TiO_2$-$NaCl_{(aq)}$ interface is displayed by the green line in Fig. 3a. Within the $TiO_2$ region, the potential exhibits pronounced oscillations, while the homogeneous and isotropic nature of the liquid electrolyte results in a more uniform profile. Macroscopic averaging of $\phi(z)$ gives the black line in Fig. 3a, from which the potential drop $\psi$, defined as the potential difference between solid and liquid phases, is extracted. At this point, the capacitance can be calculated. Taking the neutral $TiO_2$-$NaCl_{(aq)}$ interface as the reference and using the results in Fig. 1c, we have $\Delta\sigma_b = -7.54 \pm 0.13\ \mu C/cm^2$ for the TiO$_2$-NaCl+NaOH basic solution interface, and $\Delta\sigma_a = 7.69 \pm 0.04\ \mu C/cm^2$ for the TiO$_2$-NaCl+HCl acidic solution interface. By comparing the macroscopically averaged[48] electrostatic potential at the acidic and basic solution interfaces with the neutral reference (Fig. 3b), we determine the potential drop differences $\Delta\psi_a = 131.60 \pm 10.58\ mV$ and $\Delta\psi_b = -78.34 \pm 11.18\ mV$, respectively. Consequently, the differential capacitances are $C_a = 58.43 \pm 4.71\ \mu F/cm^2$ and $C_b = 96.25 \pm 13.84\ \mu F/cm^2$ under acidic and basic conditions, respectively. $C_b$ is larger than $C_a$ with $C_b/C_a = 1.6 \pm 0.3$, which agrees well with the experimental results of $C_b/C_a \approx 1.5$ at a similar interface[40] (see Supplementary Section 6 for a detailed discussion).

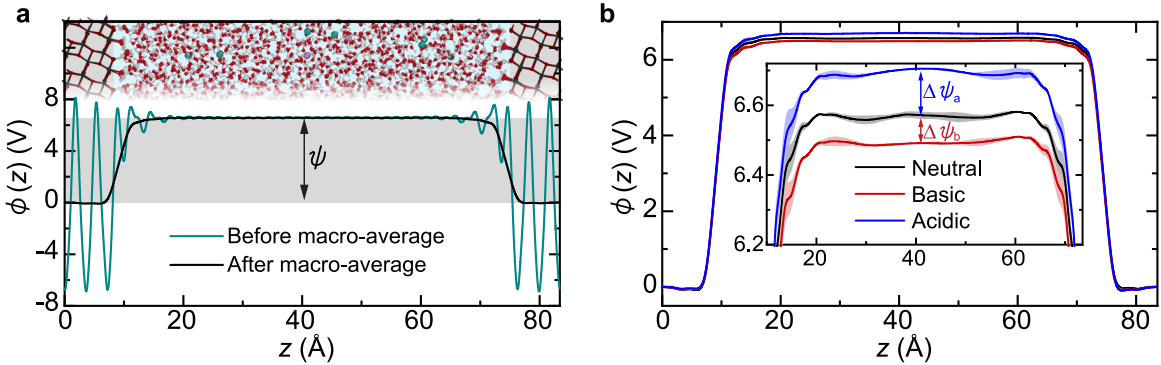

**Fig. 3 | Potential drop at interfaces. a,** Representative snapshot from our DPLR simulation of the TiO$_2$-NaCl solution interface (top) and plane-averaged electrostatic potential $\phi$ along the $z$-direction, before and after macroscopic (macro-) average (bottom). The potential drop $\psi$ is determined from the macro-averaged potential difference between the bulk solid and bulk liquid regions. **b,** Macro-averaged electrostatic potential at the TiO$_2$ interfaces with three different types of electrolytes. All curves are aligned to zero at $z = 0$. The magnified view in the inset shows the potential drop differences between the negatively (positively) charged surface and the neutral surface, denoted as $\Delta\psi_b$ ($\Delta\psi_a$), which are obtained by further averaging the macro-averaged electrostatic potentials over 31.8 Å < z < 51.8 Å. All results and error bars were derived from simulations using two independent DPLR models.

The larger capacitance observed under basic conditions in comparison to acidic ones can be understood as follows. For the negative (positive) surface, the positive Na$^+$ (negative Cl$^-$) ions are drawn towards the surface to screen the surface charge. However, the abilities of Na$^+$ and Cl$^-$ to screen the surface charge differ. Fig. 2b shows that Na$^+$ can approach the surface more closely than Cl$^-$ ions because the outermost layer of TiO$_2$ surface is composed of electronegative O$_{2c}$ atoms. This allows Na$^+$ to screen the surface charge more effectively than Cl$^-$. As revealed by Fig. 2b, transitioning from a neutral to a negative surface significantly amplifies the intensity of the Na$^+$ peak. In contrast, the transition from a neutral to a positive surface induces a modest increase in the Cl⁻ peak intensity. This leads to a smaller absolute value of $\Delta\psi_b$ than $\Delta\psi_a$, and consequently a larger $C_b$ than $C_a$. The observation that cations approach the metal oxide surface more closely than anions aligns with previous classical mean-field theory studies[40], classical force field simulations[14], and DFT calculations[43].

**Microscopic surface charging mechanism**

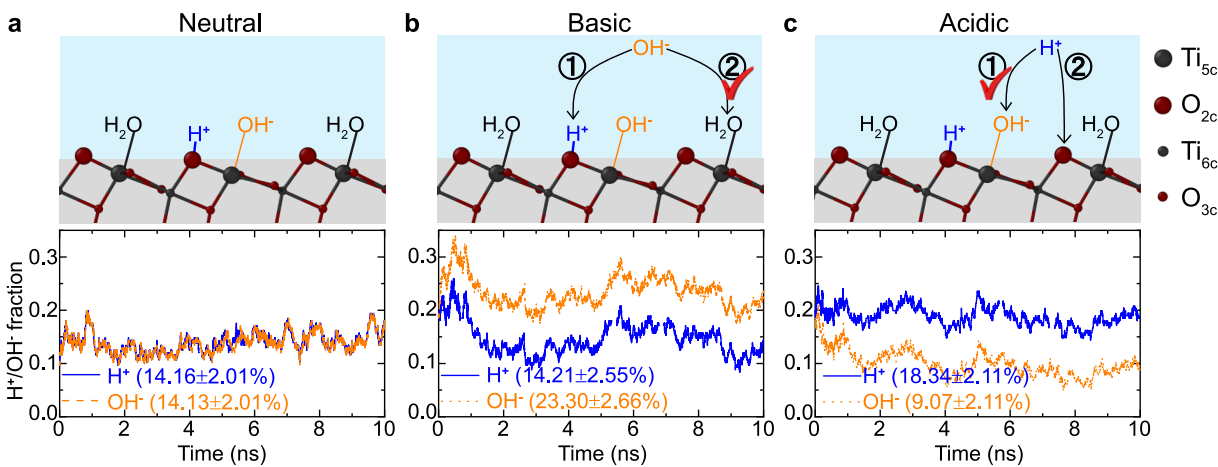

**Fig. 4 | Surface charging mechanisms.** Time evolution of the surface H⁺ and OH⁻ coverage on $TiO_2$ interfaces with **a,** neutral, **b,** basic, and **c,** acidic solutions. The legends list the surface ion coverages averaged between 3-10 ns. Schematics in **a** show molecularly absorbed $H_2O$ at $Ti_{5C}$ sites and an $H_2O$ dissociated into adsorbed H⁺ at $O_{2c}$ and OH⁻ at $Ti_{5C}$ sites. Schematics in **b** and **c** illustrate two potential pathways an additional water ion might take upon adsorption to the surface.

The distinct interfacial capacitances under acidic and basic conditions lead to different microscopic surface charging mechanisms. As shown in Fig. 4a, for the charge-neutral $TiO_2$-$NaCl_{(aq)}$ interface, 14.16 ± 2.01% water molecules adsorbed at $Ti_{5c}$ sites are dissociated into adsorbed H⁺ at $O_{2c}$ and OH⁻ at $Ti_{5c}$, with zero net surface charge. Starting with an equilibrated $TiO_2$-$NaCl_{(aq)}$ configuration, when we add bases or acids into the electrolyte, the OH⁻ or $H_3O^+$ ions have two potential pathways to be adsorbed at the surface. For an electrolyte OH⁻ (Fig. 4b), one pathway is to recombine with a surface H⁺ ion at the $O_{2c}$ site to form a water molecule in the liquid, which decreases the surface's H⁺ population. Another pathway is to replace an $H_2O$ molecule at the $Ti_{5c}$ site, which increases the number of adsorbed OH⁻ ions. Although both pathways result in the same net surface charge, the total number of surface charges, $N_{SC} = N_{H^+} + N_{OH^-}$, is different. The first pathway reduces $N_{SC}$, whereas the second increases it. Similarly, an electrolyte H⁺ can either recombine with a surface terminal OH⁻ into a water molecule or occupy an empty $O_{2c}$ site (Fig. 4c). The averaged ion coverages in Fig. 4 suggest that OH⁻ ions prefer the second pathway (increasing $N_{SC}$), whereas protons slightly prefer the first (decreasing $N_{SC}$). As a result, the average water dissociation fraction decreases significantly (to ~ 9%) with the acidic electrolyte whereas it remains approximately the same as that of the neutral interface under basic conditions. These trends can be related to the different capacitances under basic and acidic conditions, i.e. the larger capacitance at high pH permits a larger number of surface ions, whereas the opposite is true at low pH. The fact that negatively charged surfaces allow a larger density of surface hydroxyl groups is important in photocatalysis because these groups can trap photo-generated holes and form hydroxyl radicals[37,49], which are key intermediates of many photo-oxidation reactions. The photooxidation of water is indeed known to be faster at high pH[50].

**CONCLUSIONS**

The *ab initio*-level molecular-scale picture of the EDL at the TiO$_2$-electrolyte interface provided by our simulations exhibits Stern layer features that are not included in the GCS model, particularly intermediate Helmholtz planes associated with outer-sphere complexes of electrolyte ions. The shorter adsorption distance of positive vs. negative ions results in a larger capacitance at high pH, in good agreement with experiment. The different capacitances at low and high pHs are associated with distinct microscopic surface charging mechanisms, a feature that can help optimize photo(electro)catalytic processes. These results underscore the utility of machine learning-enabled, *ab initio*-quality simulations to characterize the specific chemistry and inhomogeneity of (photo)electrochemical interfaces and pave the way to further studies including the presence of external fields.

24	Behler, J. Four generations of high-dimensional neural network potentials. *Chemical Reviews* **121**, 10037-10072 (2021). https://doi.org/10.1021/acs.chemrev.0c00868

25	Unke, O. T. *et al.* Machine learning force fields. *Chemical Reviews* **121**, 10142-10186 (2021). https://doi.org/10.1021/acs.chemrev.0c01111

26	Quaranta, V., Behler, J. & Hellström, M. Structure and dynamics of the liquid–water/zinc-oxide interface from machine learning potential simulations. *The Journal of Physical Chemistry C* **123**, 1293-1304 (2019). https://doi.org/10.1021/acs.jpcc.8b10781

27	Galib, M. & Limmer, D. T. Reactive uptake of N2O5 by atmospheric aerosol is dominated by interfacial processes. *Science* **371**, 921-925 (2021). https://doi.org/doi:10.1126/science.abd7716

28	Schran, C. *et al.* Machine learning potentials for complex aqueous systems made simple. *Proceedings of the National Academy of Sciences* **118**, e2110077118 (2021). https://doi.org/doi:10.1073/pnas.2110077118

29	Zhang, C., Yue, S., Panagiotopoulos, A. Z., Klein, M. L. & Wu, X. Dissolving salt is not equivalent to applying a pressure on water. *Nature Communications* **13**, 822 (2022). https://doi.org/10.1038/s41467-022-28538-8

30	Wen, B., Calegari Andrade, M. F., Liu, L.-M. & Selloni, A. Water dissociation at the wate-rutile $TiO_2$(110) interface from ab-initio-based deep neural network simulations. *Proceedings of the National Academy of Sciences* **120**, e2212250120 (2023). https://doi.org/doi:10.1073/pnas.2212250120

31	Zeng, Z. *et al.* Mechanistic insight on water dissociation on pristine low-index $TiO_2$ surfaces from machine learning molecular dynamics simulations. *Nature Communications* **14**, 6131 (2023). https://doi.org/10.1038/s41467-023-41865-8

32	Zhang, C., Yue, S., Panagiotopoulos, A. Z., Klein, M. L. & Wu, X. Why dissolving salt in water decreases its dielectric permittivity. *Physical Review Letters* **131**, 076801 (2023). https://doi.org/10.1103/PhysRevLett.131.076801

33	Liu, R. *et al.* Structural and dynamic properties of solvated hydroxide and hydronium ions in water from *ab initio* modeling. *The Journal of Chemical Physics* **157**, 024503 (2022). https://doi.org/10.1063/5.0094944

34	Zhang, L. *et al.* A deep potential model with long-range electrostatic interactions. *The Journal of Chemical Physics* **156**, 124107 (2022). https://doi.org/10.1063/5.0083669
14

**Acknowledgements**

This work was conducted within the "Chemistry in Solution and at Interfaces" (CSI) Center funded by the USA Department of Energy under Award DE-SC0019394. The work at the Lawrence Livermore National Laboratory was performed under the auspices of the U.S. Department of Energy under Contract DE-AC52-07NA27344. This research used resources of the National Energy Research Scientific Computing Center (NERSC), a U.S. Department of Energy Office of Science User Facility located at Lawrence Berkeley National Laboratory, operated under Contract

No. DE-AC02-05CH11231. This research used resources of the Oak Ridge Leadership Computing Facility at the Oak Ridge National Laboratory, which is supported by the Office of Science of the U.S. Department of Energy under Contract No. DE-AC05-00OR22725. This research used the Princeton Research Computing resources at Princeton University which is consortium of groups led by the Princeton Institute for Computational Science and Engineering (PICSciE) and Office of Information Technology's Research Computing.

# Supplementary Information
# Electrical double layer and capacitance of TiO$_2$ electrolyte interfaces from first principles simulations

Chunyi Zhang[1], Marcos Calegari Andrade[2], Zachary K. Goldsmith[1], Abhinav S. Raman[1], Yifan Li[1], Pablo Piaggi[1], Xifan Wu[3], Roberto Car[1]*, and Annabella Selloni[1]*

[1] *Department of Chemistry, Princeton University, Princeton, New Jersey 08544, USA*

[2] *Materials Science Division, Lawrence Livermore National Laboratory, Livermore, California 94550, USA*

[3] *Department of Physics, Temple University, Philadelphia, Pennsylvania 19122, USA*

Email: rcar@princeton.edu, aselloni@princeton.edu

## Table of Contents

# 1. Deep neural network (DNN) models

## 1.1 Training dataset

The training dataset for the DNN models was collected through an active machine learning approach[1]. This dataset comprehensively spans the configurational space of bulk anatase $TiO_2$, water, and various aqueous electrolyte solutions (NaCl, NaOH, HCl, NaCl + NaOH, and NaCl + HCl solutions), as well as anatase (101) interfaces with each of these liquids (see Supplementary Table 1). The exploration spanned temperatures of 200–800 K (systems 1-11) or 300–400 K (systems 12-15), under conditions of either a pressure of 1 bar or a constant volume corresponding to experimental densities. The final dataset comprises 30,103 configurations in total.

**Supplementary Table 1**. Details of the training dataset: systems, constituents, and number of configurations ($N$) for each system.

| Systems | Constituents | $N$ |
|---|---|---|
| 1. One water molecule in a vacuum | 1 $H_2O$ | 127 |
| 2. Bulk water | 64 $H_2O$ | 2,870 |
| 3. Bulk $NaCl_{(aq)}$ | 64 $H_2O$ + 1~6 NaCl | 5,710 |
| 4. Bulk $NaOH_{(aq)}$ | 40~63 $H_2O$ + 1~12 NaOH | 4,466 |
| 5. Bulk $HCl_{(aq)}$ | 44~64 $H_2O$ + 1~12 HCl | 3,244 |
| 6. Bulk $TiO_2$ | 36 $TiO_2$ | 1,119 |
| 7. Gas-phase water on anatase (101) | 72 $TiO_2$ + 1~2 $H_2O$ | 192 |
| 8. Anatase (101)-liquid water interface | 60 $TiO_2$ + 82 $H_2O$ | 3,331 |
| 9. Anatase (101)-$NaCl_{(aq)}$ interface | 60 $TiO_2$ + 82 $H_2O$ + 1 NaCl | 3,794 |
| 10. Anatase (101)-$NaOH_{(aq)}$ interface | 60 $TiO_2$ + 60~81 $H_2O$ + 1~12 NaOH | 2,590 |
| 11. Anatase (101)-$HCl_{(aq)}$ interface | 60 $TiO_2$ + 71~82 $H_2O$ + 1~12 HCl | 1,718 |
| 12. Bulk $NaCl_{(aq)}$ +$NaOH_{(aq)}$ | 61 $H_2O$ + 1 NaCl + 1 NaOH | 147 |
| 13. Bulk $NaCl_{(aq)}$ +$HCl_{(aq)}$ | 61 $H_2O$ + 1 NaCl + 1 HCl | 152 |
| 14. Anatase (101)-$NaCl_{(aq)}$+$NaOH_{(aq)}$ interface | 60 $TiO_2$ + 74~80 $H_2O$ + 1~2 NaCl + 1~4 NaOH | 346 |
| 15. Anatase (101)-$NaCl_{(aq)}$+ $HCl_{(aq)}$ interface | 60 $TiO_2$ + 78~81 $H_2O$ + 1~2 NaCl + 1~4 HCl | 297 |

**1.2 Testing dataset**

To evaluate the performance of our DNN models, we generated a testing dataset of configurations not included in the training. This was achieved by conducting deep potential long-range (DPLR) molecular dynamics (MD) simulations on the following four representative systems.

- The anatase (101)-liquid water interface, comprising 60 $TiO_2$ units and 82 $H_2O$ molecules.
- The anatase (101)-$NaCl_{(aq)}$ solution interface, comprising 60 $TiO_2$ units, 82 $H_2O$ molecules, and 1 NaCl ion pair.
- The anatase (101) -$NaCl_{(aq)}$+$NaOH_{(aq)}$ solution interface, comprising 60 $TiO_2$ units, 80 $H_2O$ molecules, 1 NaCl ion pair, and 1 NaOH ion pair.
- The anatase (101)-$NaCl_{(aq)}$+ $HCl_{(aq)}$ solution interface, comprising 60 $TiO_2$ units, 81 $H_2O$, 1 NaCl ion pair, and 1 HCl ion pair.

In the above, the 60 $TiO_2$ units correspond to a five-layer (1 × 3) anatase (101) slab. For each system, we conducted 5 ns DPLR MD simulations within the canonical ensemble at 330 K. The initial 1 ns of each simulation was discarded for equilibration purposes. From the subsequent 4 ns, 50 configurations were uniformly extracted from each trajectory, resulting in a total of 200 configurations for the testing dataset.

**1.3 *Ab initio* calculations**

Total energies and atomic forces for the training and testing datasets were calculated within density functional theory (DFT)[2] using the strongly constrained and appropriately normed (SCAN)[3] exchange-correlation functional as implemented in the Quantum ESPRESSO[4] package. The SCAN functional[3] has been found to well describe $TiO_2$ interfaces[5] and electrolyte solutions[6-8] in previous studies. The electron-nuclei interactions were described by Optimized Norm-Conserving Vanderbilt (ONCV)[9] pseudopotentials. Electron wavefunctions were expanded in plane waves using a cutoff energy of 150 Ry. A total energy convergence threshold of $1 \times 10^{-6}$ Ry was adopted. Because of the large size of our supercells, the Brillouin zone sampling was restricted to the Gamma point. Following each self-consistent calculation, maximally localized Wannier functions (MLWFs)[10] were determined using the Wannier90 code[11]. Each MLWF was associated with its closest Ti, O, Na, or Cl atom, resulting in each of these atoms carrying four doubly occupied MLWFs. From the MLWFs, the coordinates of the Wannier Centroids (WCs) relative to their corresponding atoms were obtained by computing the average position of the Wannier centers associated with each given atom. The DPLR and Deep Wannier (DW)[12] DNN models were trained on the set of DFT-SCAN training data using the DeePMD-kit package[13]. For both DPLR and DW, two independent models were generated using different initial random parameters. The two models were used to run independent simulations from which average properties and corresponding error bars were derived.

**1.4 DPLR method**

The DPLR method[14] assumes that the potential energy surface has short- and long-range contributions. The short-range contribution is represented as in the standard deep potential model[15], while the long-range contribution is approximated by the electrostatic energy of a system of spherical Gaussian charges associated with the ions (nuclei + core electrons) and the valence electrons. We calculate the electrostatic energy of the Gaussians via the particle-particle-particle-mesh method[16] for evaluating Ewald sums. The location, charge magnitude, and spatial spread of the Gaussians are determined as follows.

- **Location:** Ionic Gaussians are centered at the atomic sites, and electronic Gaussians are centered at the maximally localized Wannier centers[17]. For computational simplicity, Wannier centers associated with the same atom are combined into a single Wannier Centroid (WC) [14], located at the instantaneous average position of these Wannier centers. The locations of WCs are predicted by the DW DNN model[12].

- **Charge magnitude:** The magnitudes of the ionic charges are $+Z_V e$, with $Z_V$ being the number of their valence electrons. The pseudopotential applied in this work treats the $3s^2 3p^6 3d^2 4s^2$ electrons of Ti, 1s electron of H, $2s^2 2p^4$ electrons of O, $2s^2 2p^6 3s^1$ electrons of Na, and $3s^2 3p^5$ electrons of Cl as valence electrons explicitly. Consequently, $Z_V$ equals 12, 1, 6, 9, and 7 for Ti, H, O, Na, and Cl, respectively. For electrons, each Wannier center carries a charge of $-2e$. Given that each Ti, O, Na, and Cl ion has four Wannier centers, their WCs carry a charge of $-8e$. The DFT calculations show that the average distance of Ti's WCs from the Ti ions is significantly smaller (0.002 Å, averaged over $x$, $y$, and $z$ directions) compared to that of O's WCs from O ions (0.026 Å, averaged over $x$, $y$, and $z$ directions). This suggests a negligible contribution of the polarization of Ti's WCs to the overall electrostatic energy. Therefore, for computational efficiency, our DPLR model further simplifies the representation of Ti ions: instead of accounting for a $+12e$ charge on the Ti ion counterbalanced by a $-8e$ charge from its WCs, we simplify our DPLR model by omitting the WCs of Ti and treating each Ti ion as a $+4e$ charge.

- **Spatial spread:** The long-range electrostatic contribution up to dipole terms is independent of the spread parameter, which can then be chosen to ensure charge neutrality in the bulk of the solution as well as good numerical representability by a DNN of the short-range contribution. In our DPLR, the spatial spread of the Gaussians for both ions and electrons is $(2\beta)^{-1}$, where $\beta$ is an adjustable parameter. In the limit of $\beta \to 0$, the Gaussian width is infinite and the DPLR model reduces to the standard short-range DP (hereafter denoted as DPSR) model. Conversely, as $\beta \to \infty$, the Gaussian charges become point-like, leading to singular potentials incompatible with DNN representations.[14] To determine an optimal $\beta$ for our system, we conducted systematic training of DPLR models across a range of $\beta$ values: 0.0, 0.1, 0.2, 0.3, and 0.4 Å$^{-1}$. The performance analysis, depicted in Supplementary Fig. 1, reveals that, for both the training and testing datasets, $\beta = 0.1$ Å$^{-1}$ minimizes the root mean square error (RMSE) in the prediction of energies and forces in comparison to the DFT calculations. We thus adopted $\beta = 0.1$ Å$^{-1}$ for our DPLR simulations. With this choice, charge neutrality in the bulk is insured (see Supplementary Figure 5), and the short-range description of energy and forces is at least as good as, in fact slightly better than, DPSR, as shown below (Supplementary Figure 1).

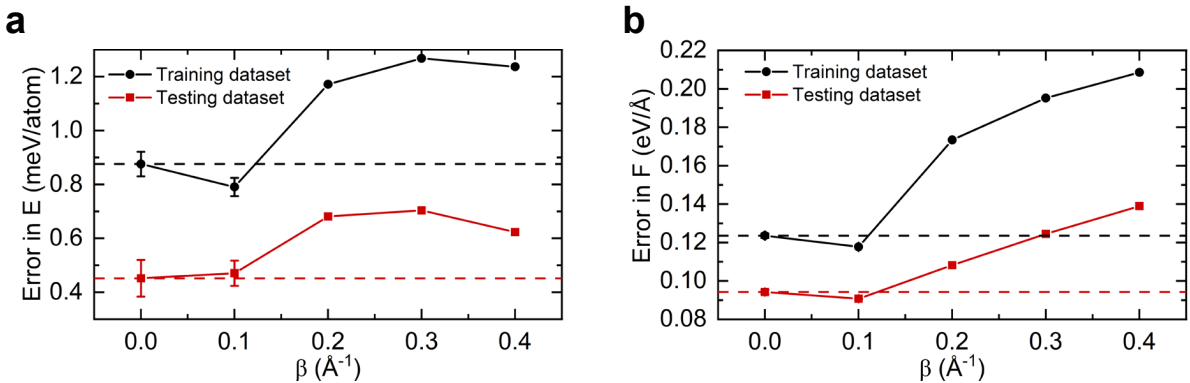

**Supplementary Figure 1.** Root mean square error of the DPLR-predicted **a**, energy, and **b,** atomic force as a function of the spread parameter $\beta$, evaluated on training and testing datasets. The horizontal dashed lines indicate the errors predicted by DPSR.

Supplementary Fig. 1 also shows that the RMSE for the testing dataset is lower than that for the training dataset. This is because the training dataset includes higher temperature (up to 800 K) configurations compared to the 330 K temperature used for the testing dataset. At higher temperatures, the magnitudes of forces and energies are larger, resulting in larger absolute errors. Given that the production runs of our work were conducted at 330 K, the errors at 330 K are more representative and indicative of the model's performance under our specific conditions of interest.

**1.5 DPLR MD simulations**

The large-scale DPLR MD simulations presented in the main manuscript were conducted on model systems consisting of a five-layer (3 × 9) anatase (101) slab (540 $TiO_2$ units) in contact with a 67 Å thick layer of aqueous electrolyte within a periodically repeated supercell of size 30.7 Å × 33.9 Å × 83.4 Å along the three orthogonal directions $[\bar{1}01]$, $[010]$ and $[101]$ of the anatase crystal lattice. All simulations were conducted in the canonical ensemble for 10 ns with a temperature of 330 K. The 30 K elevation is to partially compensate for the overestimation of the melting temperature of ice by the SCAN functional and describe a liquid with diffusivity close to that of water at standard conditions[18]. The $TiO_2$ interfaces with the acidic and basic solutions were initially simulated at 400 K for 0.9 ns to accelerate equilibration and subsequently cooled to the target temperature of 330 K.

**1.6 Validation of the DNN models**

1.6.1 Energies, atomic forces, and WCs

We assessed the performance of our DPLR and DW DNN models by comparing their predicted energies, atomic forces, and WCs for the testing dataset to the results of DFT-SCAN calculations. Supplementary Fig. 2 shows that the DNN models reproduce well the DFT results. The root-mean-squared errors of the energies, atomic forces, and WCs predicted by the DNN models with respect to DFT are 0.47 meV/atom, 0.091 eV/Å, and 0.003 Å, respectively.

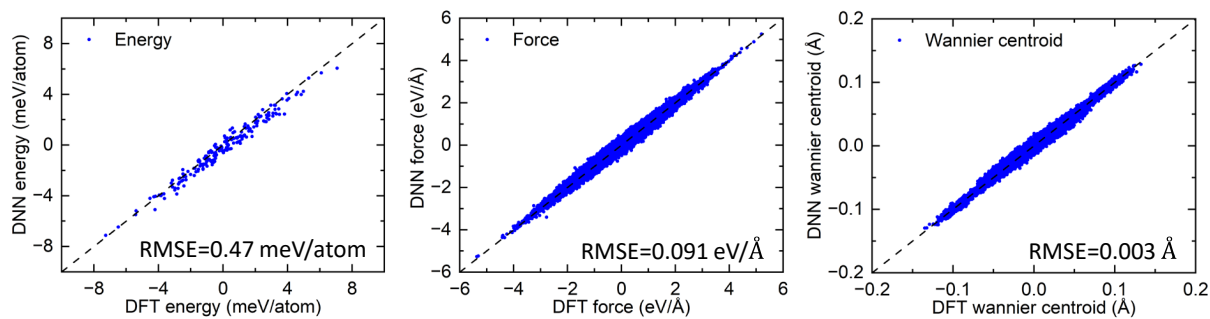

**Supplementary Figure 2.** Comparison between the energies, atomic forces, and Wannier centroids predicted by the DNN models and DFT-SCAN calculations. The average value of the energy of each system was shifted to zero for better visualization.

1.6.2 Radial distribution functions and water density profiles

We also compared the radial distribution functions of selected atomic type pairs and water density profiles at the anatase (101)-water interface predicted by DPLR molecular dynamics (DPLR MD) with DPSR molecular dynamics (DPSR MD) and previously reported results from *ab initio* molecular dynamics (AIMD)[5]. As shown in Supplementary Fig. 3, the agreement between DPLR-MD, DPSR-MD, and AIMD simulations (both on the time scale of ~ 40 ps) is quite satisfactory.

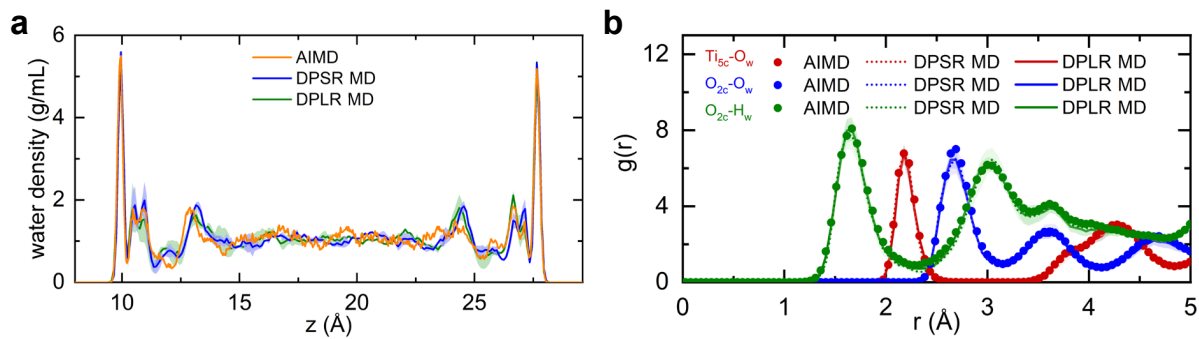

**Supplementary Figure 3.** Comparison between the predictions of DPLR MD, DPSR MD, and AIMD[5] simulations (both on a time scale of ~ 40 ps) for the water-anatase (101) interface: **a,** Water density profile along the direction perpendicular to the $TiO_2$ surface; **b,** radial distribution functions, $g(r)$, of selected atomic type pairs. The definition of $Ti_{5c}$ and $O_{2c}$ is given in the main manuscript, and $O_w$ denotes the water oxygen atoms. Shaded areas indicate the standard deviation obtained from two independent DPLR MD or two independent DPSR MD simulations.

1.6.3 Interfacial water dissociation and potential of mean force

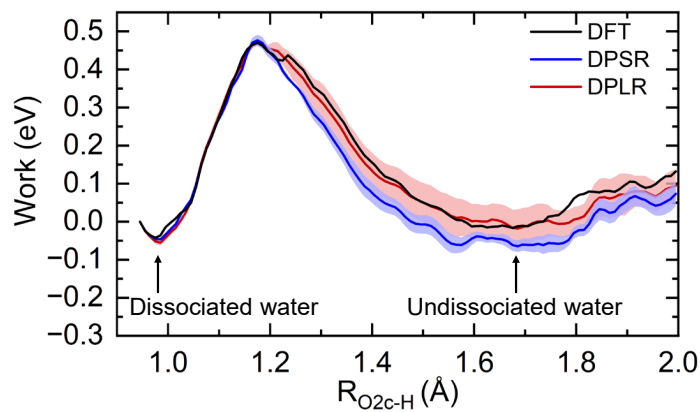

**Supplementary Figure 4.** Comparison between DFT, DPSR, and DPLR results for the work to move an $H^+$ from a surface $O_{2C}$ to an $OH^-$ adsorbed on an adjacent $Ti_{5c}$ at the anatase (101)-water interface. The two local minima correspond to dissociated and undissociated water. Shaded areas indicate the standard deviation obtained from two independent DNN models.

Following Ref. 19, we computed the work required to move an $H^+$ from a surface $O_{2C}$ to an $OH^-$ adsorbed on an adjacent $Ti_{5c}$ at the anatase (101)-water interface (see below for computational details). Supplementary Fig. 4 shows the results obtained using DFT, DPSR, and DPLR. DPLR is in closer agreement with DFT than DPSR. The higher energy of undissociated water predicted by DPLR in comparison to DPSR is in accordance with the larger water dissociation fraction observed in DPLR-MD simulations (14.1 ± 2.0%) relative to that reported by previous DPSR simulations (5.6 ± 0.5%)[19].

*Computational details.* To compute the curves in Supplementary Fig. 4, we conducted a 2.5 ns enhanced sampling simulation with only one of our two DPLRs (referred to as DPLR model1), and the $TiO_2$-water interface was modeled as a (1 × 3) anatase (101) slab in contact with a 20 Å slab of water. Using a reaction coordinate defined as the minimum distance between a particular surface $O_{2C}$ atom and any H atom in the system (denoted $R_{O2c-H}$), we applied an external bias potential designed to enhance the water dissociation at the interface. We extracted atomic configurations with different $R_{O2c-H}$ values from the simulation trajectory and calculated the force projected on the unit vector connecting an $O_{2C}$ to the nearest H atom of these configurations using two DPLR models ($f_{\text{DPLR model}j}$, $j$=1, 2), as well as two DPSR models ($f_{\text{DPSR model}j}$), and DFT ($f_{\text{DFT}}$).

Since the configurational space was explored using DPLR model1, a reweighing process was necessary for the other models. The reweighted force was performed as follows:

$$f_*^{reweighted} = \frac{<f_* \times e^{\frac{E_{\text{DPLR model1}}-E_*}{k_BT}}>_{\text{DPLR model1}}}{<e^{\frac{(E_{\text{DPLR model1}}-E_*)}{k_BT}}>_{\text{DPLR model1}}}, \quad (1)$$

where * could be any of the two DPLR models, two DPSR models, or DFT. The notation $<\cdots>_{\text{DPLR model1}}$ indicates that the average is conducted on the configurations extracted from the DPLR model1 trajectory, $k_B$ is the Boltzmann constant, $T$ is the simulation temperature which equals to 330 K, and $E_*$ is the energy of the system predicted by model *. The work associated with proton transfer was then computed as the integral of reweighted forces.

## 2. Comparing the ion distributions from DPSR and DPLR simulations

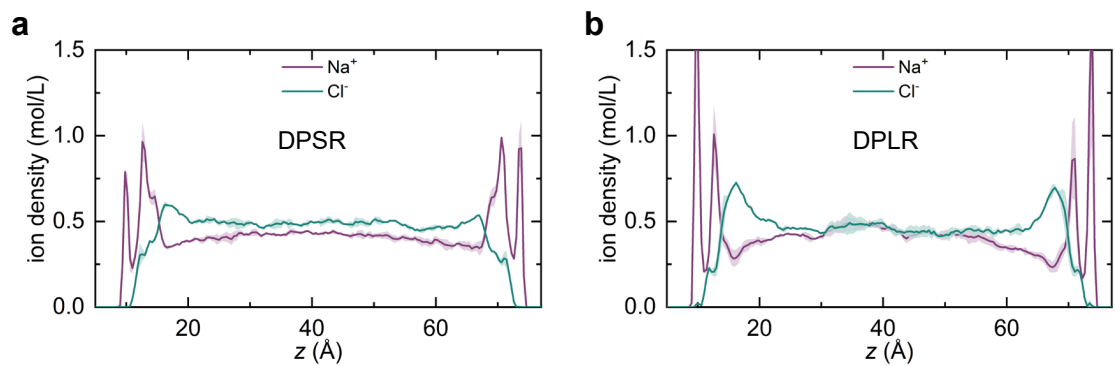

**Supplementary Figure 5.** Plane-averaged ion distributions along the $z$-direction for the TiO$_2$-NaCl solution interface obtained from DPSR and DPLR MD simulations. The 5 ns simulations were conducted within the canonical ensemble at 600 K, rather than 330 K, to leverage the faster statistical convergence achievable at elevated temperatures.

Supplementary Fig. 5 shows the ion distributions at 600 K obtained from DPSR and DPLR MD simulations. Due to the screening effect of the EDL, the solution should recover its bulk properties in the central region (away from the interface) and thus have equal densities of Na$^+$ and Cl$^-$ ions. DPSR simulations predict instead a ~ 0.06 M higher density of Cl$^-$ vs Na$^+$ ions in the central region (Supplementary Fig. 5a).   Because of the lack of long-range electrostatic interactions, there is no energy penalty for the unphysical charge imbalance in the DPSR models. The inclusion of long-range electrostatic interactions corrects this issue, as shown by the DPLR results in Supplementary Fig. 5b. The ion distributions within the EDL are also changed after including these long-range interactions, confirming that these are essential for the correct description of the EDL.

## 3. PH point of zero proton charge (pH$_{PZC}$)

As a reference for the surface charging results reported in the main text (Fig. 1), we estimated the pH$_{PZC}$ of the aqueous anatase (101) surface using our DPLR model. The pH$_{PZC}$ of an oxide surface is determined by the acid dissociation constants ($pK_a$s) of the surface acid-base active sites that can accept or release protons. On anatase (101), there are two types of such sites, the O$_{2c}$ and the Ti$_{5c}$ sites, and the pH$_{PZC}$ is simply the average of the $pK_a$s of these sites, i.e., pH$_{PZC}$ = $(pK_{a,O_{2c}} + pK_{a,Ti_{5c}})/2$.

We evaluated these $pK_a$s from the free energy changes associated with the following reactions:

**reaction 1** (for the O$_{2c}$ site)

$$O_{2c}H^+ + H_2O_{(aq)} \rightarrow O_{2c} + OH_3^+{}_{(aq)} \quad (2)$$

**reaction 2** (for the Ti$_{5c}$ site)

$$Ti_{5c}H_2O + H_2O_{(aq)} \rightarrow Ti_{5c}OH^- + H_3O^+_{(aq)} \quad (3)$$

This study was performed via DPLR MD and enhanced sampling (well-tempered metadynamics) simulations using the LAMMPS package[20] with the DeepMD-kit[13] and PLUMED[21]. For each reaction, we carried out two independent DPLR simulations, from which we derived the average properties and their error bars. The simulations were performed in the canonical ensemble at 330 K for 2ns on a system comprised of a (1 × 3) anatase (101) slab (exposing 12 O$_{2c}$ and Ti$_{5c}$ sites) in contact with a 20 Å thick layer of water. For the enhanced sampling, we used the collective variables (CVs) introduced in Ref. 22 and followed the procedure described in Ref. 23. For reaction 1, the reference state had 11 undissociated water molecules adsorbed at 11 Ti$_{5c}$ sites and one dissociated water molecule with OH$^-$ adsorbed at a Ti$_{5c}$ site and H$^+$ adsorbed at a nearby O$_{2c}$ site. We applied restraints to only allow proton transfer between the specific O$_{2c}$ site and liquid water. For reaction 2, the reference state had 12 undissociated water molecules adsorbed at 12 Ti$_{5c}$ sites. Here, restraints were introduced to permit only one of the 12 water molecules to exchange one proton with liquid water at a time. These constraints were essential for the successful implementation of the CVs[22,23].

Supplementary Fig. 6 shows the free energy surface of the two reactions as a function of the $s_p$ and $s_d$ CVs defined in Refs. 22,23. For each reaction, the deprotonation free energy, $\Delta F$, was calculated as the free energy difference between the deprotonated state ($s_p \approx 1$, $10\,\text{Å} < s_d < 14\,\text{Å}$) and the protonated state ($s_p \approx 0$, $s_d \approx 0$). For the deprotonated state, the range $10\,\text{Å} < s_d < 14\,\text{Å}$ ensures that the released proton is sufficiently distanced from both the upper and lower interfaces, preventing interfacial effects on the free proton. In this way, $\Delta F$ for reactions 1 and 2 were calculated to be $37.25 \pm 0.62$ and $51.00 \pm 0.65$ kJ/mol, respectively. Using $pK_a = \Delta F/2.303 k_B T$, we obtained $pK_{a,O_{2c}} = 5.9 \pm 0.1$ and $pK_{a,Ti_{5c}} = 8.1 \pm 0.1$, yielding pH$_{\text{PZC}} = 7.0 \pm 0.1$. This result compares well to the experimental pH$_{\text{PZC}}$ range of $6 \pm 1$[24-26].

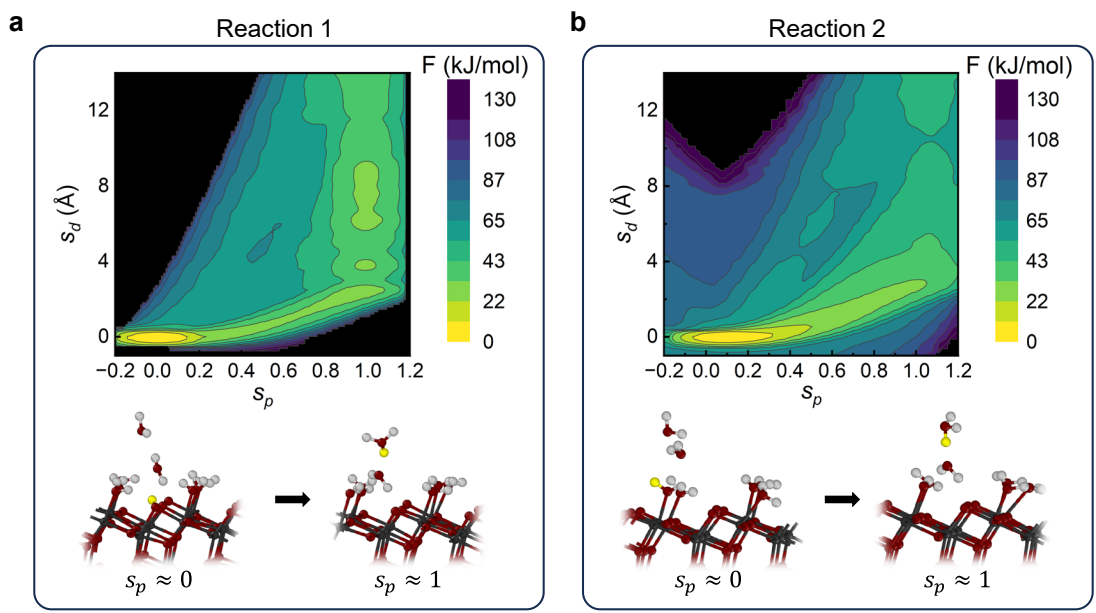

**Supplementary Figure 6.** Free energy surface for **a,** reaction 1 and **b,** reaction 2. The bottom panels show the representative snapshots of the proton transfer process from $s_p \approx 0$ to $s_p \approx 1$. For visualization purposes, only the relevant atoms are shown. Color code: Ti, grey; O, red; H, white (the proton being transferred is highlighted in yellow).

## 4. Water distribution and orientation

Supplementary Fig. 7 displays the water density distributions across the $TiO_2$ interfaces with different electrolytes. Here the 1st, 2nd, 3rd, and 4th water layers are defined in accordance with Ref. 5. We can see small differences among the water density distributions at the different interfaces, notably in the peak intensities of the 1st water layer, and the first and second sub-peaks of the 2nd water layer. These differences can be primarily attributed to the different amounts of $H^+$ and $OH^-$ ions adsorbed on the $TiO_2$ surface for the different electrolytes.

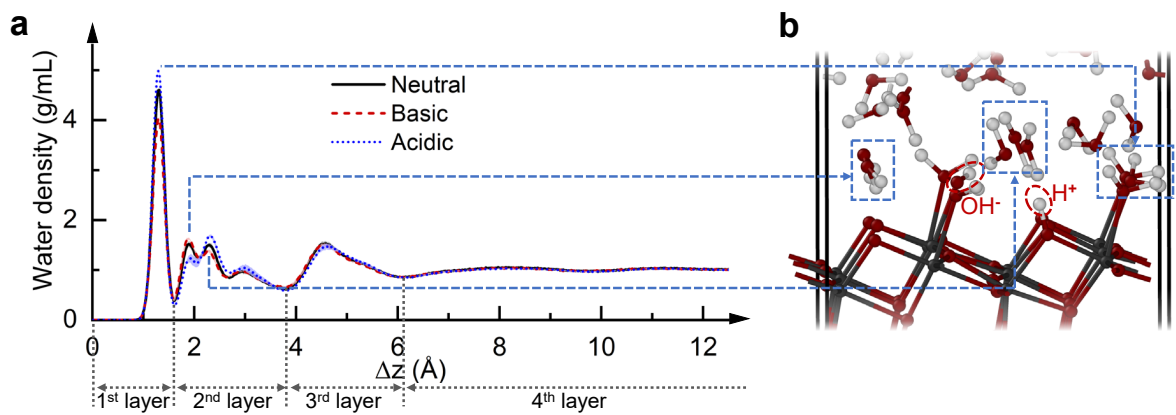

**Supplementary Figure 7. a,** Water density distributions as a function of distance, $\Delta z = z - z_{\text{surface}}$, from the solid surface across the $TiO_2$ interfaces with different electrolytes. The results are averaged over the two interfaces in the supercell, and the position of the solid surface, $z_{\text{surface}}$, is defined as the average location of the surface $O_{2c}$ atoms. **b,** A representative snapshot of the interface showing the water molecules corresponding to distinct peaks observed in the water density distribution.

As shown in Fig. 4 of the main manuscript, the $TiO_2$ surface's interaction with the basic, neutral and acidic electrolytes results in different coverage ratios of $OH^-$ on the $TiO_2$ surface, with the order being basic ($23.3 \pm 2.7\%$) > neutral ($14.1 \pm 2.0\%$) > acidic ($9.1 \pm 2.1\%$). Conversely, the coverage ratios for $H^+$ are acidic ($18.3 \pm 2.1\%$) > neutral ($14.2 \pm 2.0\%$) ≈ basic ($14.2 \pm 2.0\%$). Given that the 1st water layer is comprised of adsorbed water species at $Ti_{5c}$ sites, the occupancy of $Ti_{5c}$ sites by $OH^-$ reduces the quantity of intact water molecules in the 1st layer. Consequently, the intensity of the density peak for the 1st layer of water molecules in Supplementary Fig. 7a follows the sequence: basic < neutral < acidic.

The 2$^{nd}$ water layer comprises three sub-peaks (see Supplementary Fig. 7a). As shown in Supplementary Fig. 7b, the water in the first sub-peak (i.e., closer to the surface) generally forms two hydrogen bonds (H-bonds) with the surface $O_{2c}$ atoms. The water molecules in the second sub-peak, situated further away from the surface, typically form one H-bond with either the surface $O_{2c}$ atoms or an adsorbed $OH^-$. As shown in Supplementary Fig. 7b, when $H^+$ is adsorbed on $O_{2c}$, the 2$^{nd}$ layer water tends to form only one H-bond with the surface. Given the coverage order of $H^+$ as acidic > neutral ≈ basic, the intensity of the first sub-peak then follows the trend: acidic < neutral ≈ basic, whereas the second sub-peak's intensity follows the trend: acidic > neutral ≈ basic, as indeed shown in Supplementary Fig. 7a. Beyond the second sub-peak, the densities of water molecules in basic, neutral, and acidic systems become very similar, indicating a reduced impact of the interface on water structuring at further distances from the interface.

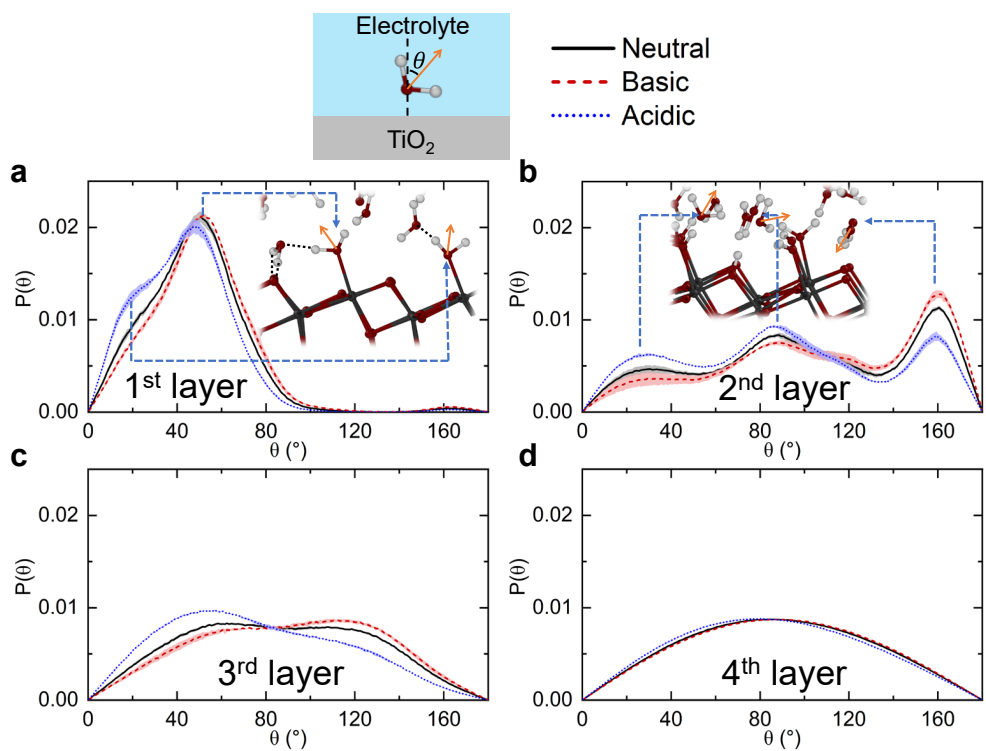

**Supplementary Figure 8.** Normalized probability distributions of the angle ($\theta$) between the unit vector bisecting the two OH groups of a water molecule (indicated by orange arrows) and the surface normal for $TiO_2$ interfaces with various electrolytes. The 1$^{st}$, 2$^{nd}$, 3$^{rd}$, and 4$^{th}$ water layers are defined in Supplementary Fig. 7a. Insets show representative atomic configurations corresponding to various peaks observed in the angular distributions. Dashed black lines show representative hydrogen bonds. For visualization purposes, only the most relevant atoms are shown.

The water angular distributions shown in Supplementary Fig. 8 reveal small differences among different electrolytes, which can also be attributed to the different amounts of $H^+$ and $OH^-$ ions adsorbed on the surface. For the water molecules in the 1$^{st}$ water layer, the O atoms are adsorbed at $Ti_{5c}$ sites while the hydrogen atoms form H-bonds with water molecules in the 2$^{nd}$ water layer, resulting in a major peak at ~50° and a shoulder at ~20° in the $\theta$ distribution as shown in Supplementary Fig. 8a. The major peak arises from water molecules in the 1$^{st}$ water layer that are H-bonded to the water molecules with two hydrogens pointing towards the surface[27], while the shoulder is contributed by water molecules in the 1$^{st}$ water layer that are H-bonded to the water molecules with only one hydrogen pointing towards the surface. When $H^+$ ions are adsorbed at the surface $O_{2c}$, the nearby water molecules in the 2$^{nd}$ water layer prefer to have only one hydrogen pointing towards the surface, explaining why the shoulder is most pronounced for the interface with the acidic solution.

For the 2$^{nd}$ water layer, the distribution of $\theta$ features three distinct peaks centered around 30°, 90°, and 160°. The representative water molecules contributing to each peak are shown in the inset of Supplementary Fig. 8b. The peak at 30° is predominantly due to water molecules that orient both hydrogen atoms away from the surface while accepting H-bonds from the 1$^{st}$ layer water. The peak at 90° arises mainly from water molecules with one H atom pointing towards the surface (either donating an H-bond to surface $O_{2c}$ or to surface $OH^-$) and the other pointing away from the surface. Lastly, the peak at 160° is mostly contributed by water molecules that direct both hydrogen atoms towards the surface and form two H-bonds with surface $O_{2c}$ atoms. As already pointed out, the presence of bridging hydroxyls (i.e., $O_{2c}$ with an adsorbed $H^+$) makes the water molecules orient with their H atoms away from the surface and donate fewer H-bonds to the surface, resulting in a small angle $\theta$. Consequently, the acidic solution, which has the highest $H^+$ surface coverage among the three electrolytes, shows a greater probability of smaller $\theta$ angles, as shown in Supplementary Fig. 8b.

The water molecules in the 3$^{rd}$ water layer, being further away from the surface, are more influenced by the net surface charge than by their interaction with specific $H^+$ or $OH^-$ groups adsorbed on the $TiO_2$ surface. This surface charge causes water molecules in a basic (acidic) electrolyte to orient their hydrogen (oxygen) atoms—towards the negatively (positively) charged surface, resulting in a larger (smaller) angle $\theta$ (see Supplementary Fig. 8c).

Regardless of the electrolyte type, the water molecules in the 4th layer exhibit a uniform distribution similar to that in bulk electrolyte solutions (Supplementary Fig. 8d). This suggests that these water molecules are almost unaffected by the surface charge because the surface charge has been well-screened by the electrical double layer.

We note that the distribution and orientation of water molecules are also influenced by the salt ions and not only by the adsorbed $H^+$ and $OH^-$ ions at the surface. As elucidated in Ref. 7, water molecules in the $Na^+$ ion's first hydration shell preferentially orient with their oxygen ends towards $Na^+$, while water molecules in the $Cl^-$ ion's first hydration shell preferentially orient the hydrogen end towards $Cl^-$. In this study, we did not conduct a quantitative analysis of this effect because the influence exerted by salt ions is very small relative to the impact of the surface charge. This relatively minor influence is attributed to two primary factors: first, the salt concentrations studied in this work are low; second, within the hydration shell of a specific ion, water molecules are oriented in diverse directions, leading to a relatively uniform distribution rather than generating pronounced peaks in the probability distributions of the angle $\theta$.

## 5. Electrostatic potential

To calculate the interfacial capacitance, we need the potential drop at the interface, and thus the total electrostatic potential along the surface normal $z$, $\phi(z)$, which is given by the sum of the contributions of the ions (nuclei + core electrons), $\phi_i(z)$, and the valence electrons, $\phi_e(z)$, namely $\phi(z) = \phi_i(z) + \phi_e(z)$. The ionic term $\phi_i(z)$ is just the potential of spherical Gaussian charges with spreads given by the pseudopotential[9] to eliminate the singularities associated to point charges. While only the long-range electrostatic energy up to dipole contributions is needed for DPLR simulations, $\phi_e(z)$ depends on the actual electron density distribution. This is well approximated by the sum of Gaussian distributions centered at the WCs (obtained from the DW DNN), with spreads given by the spherical average of the spreads calculated from DFT. Unlike in the DPLR models, in this calculation, each Ti was treated as a $+12e$ ion counterbalanced by a $-8e$ charge from its WCs. The potential $\phi_e(z)$ was then calculated by Fourier transform of the following reciprocal space expression:

$$\phi_e(m_z) = \frac{-1}{\pi m_z^2} \sum_{i=1}^{n_{WC}} q_{WC}^i \exp(-i2\pi m_z z_{WC}^i) \exp\left(-\frac{\pi^2 m_z^2}{2\beta_z^{i\,2}}\right), \qquad (4)$$

Here $m_z = \frac{K_z}{2\pi}$ where $K_z$ is a reciprocal lattice vector in the $z$ direction, $n_{WC}$ is the total number of WCs, $z_{WC}^i$ is the $z$-coordinate of the $i$th WC, $(2\beta_z^i)^{-1}$ the corresponding $z$-spread, and $q_{WC}^i$ ($=-8e$) is the charge of the $i$th WC. The negative sign before the summation indicates that $\phi_e(z)$ is the electrostatic potential experienced by electrons. Note that the term $K_z = 0$ in Eq. (4) is cancelled by the analogous term from $\phi_i(z)$.

To validate our procedure, we compared the electrostatic potential calculated using our methodology with the potential given by DFT, both averaged over 50 configurations extracted from DPLR MD simulations of a (1 × 3) anatase (101) slab interface slab in contact with a 20 Å thick layer of aqueous NaCl solution (see Supplementary Fig. 9a). As illustrated in Supplementary Fig. 9b, our computed $\phi(z)$ closely reproduces the DFT result, confirming the accuracy and reliability of our approach.

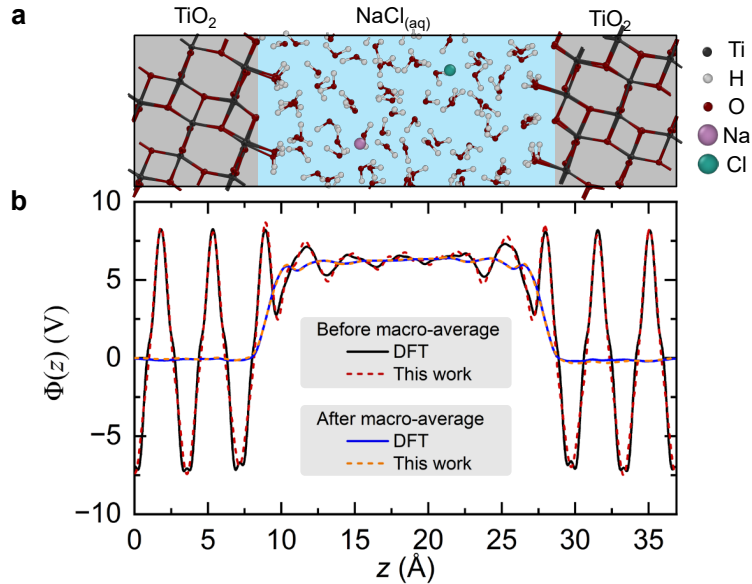

**Supplementary Figure 9. a,** Representative snapshot of the anatase (101)-NaCl solution interface used to validate our procedure for calculating the electrostatic potential. **b,** Plane-averaged electrostatic potential, $\phi(z)$, along the $z$-direction of the interface, before and after macro-average[28]. Results obtained using DFT and our developed methodology are presented in solid and dashed lines, respectively.

## 6. Finite cell size limitation and comparison to experiment

Although the size of our simulation box is quite large by AIMD standards, it is not large enough to allow establishment of equilibrium in the exchanges of water ions between the surface and the bulk reservoir. This limitation prevents us from reproducing the relationship between pH and surface charge density, $\sigma$, that is observed in experiments. Specifically, in our simulations all $H^+$ and $OH^-$ ions in the acidic and basic electrolyte were adsorbed on the $TiO_2$ surface, leading to $\sigma_a \approx 7.7\ \mu C/cm^2$ and $\sigma_b \approx -7.5\ \mu C/cm^2$, respectively, with no water ions in the bulk region of the electrolyte. In the experiment, on the other hand, such values of $\sigma_a$ and $\sigma_b$ are observed at solution pH's of $\approx$ 4.4 and 7.4[25], respectively, two pH values that are not accessible to our simulations.

However, the above limitation does not affect our results for the structure and capacitance of the EDL because a pH value of 4.4 (or 7.4) corresponds to a negligible amount of $\sim 7 \times 10^{-7}$ $H^+$ or $OH^-$ ions in our electrolyte solution. As shown in Supplementary Fig. 10, our calculated differential capacitances are only slightly smaller than the experimental values. This underestimate may in part be attributed to the finite difference approximation, $C = \frac{d\sigma}{d\psi} \approx \frac{\Delta\sigma}{\Delta\psi}$, adopted in this work because the experimental titration curve[29,30] exhibits a larger $\frac{d\sigma}{d\psi}$ than $\frac{\Delta\sigma}{\Delta\psi}$ at corresponding surface charge density. Nonetheless, our calculated ratio of $\frac{C_b}{C_a} = 1.6 \pm 0.5$ agrees well with the experimental results[29,30] of $\frac{C_b}{C_a} \approx 1.5$ at a similar interface (the rutile interface with 0.1 M $NaNO_3$ used as the background salt in Supplementary Fig. 10).

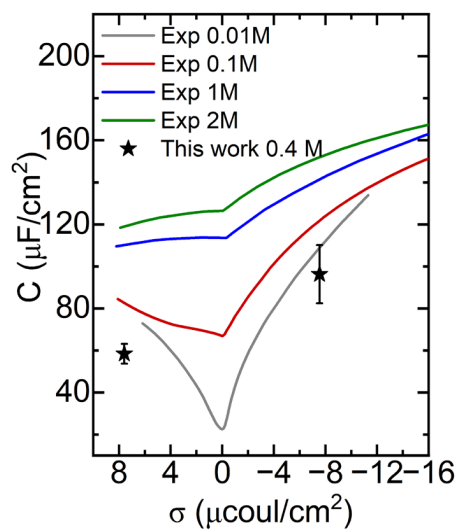

**Supplementary Figure 10.** Interfacial capacitance from experiments[29,30] (solid lines) and from the calculations of this work (star symbols). Different colors represent different concentrations of the background salt. Although the experimental results refer to rutile interfaces with $NaNO_3$ used as the background salt, whereas our simulations were conducted on anatase interfaces with NaCl as the background salt, previous experiments indicate that the surface chemistries of rutile and anatase are similar, and the difference between the effects of $NO_3^-$ and $Cl^-$ on the capacitance is negligible[29,30].